\documentclass[preprint2]{aastex}

\newcommand{\myemail}{awaki@astro.phys.sci.ehime-u.ac.jp}

\slugcomment{}

\shorttitle{A variability study of NGC 6300}
\shortauthors{Awaki et al.}

\begin{document}


\title{A variability study of the Seyfert 2 galaxy NGC 6300 with {\it XMM-Newton}}


\author{Hisamitsu Awaki, Hirokatsu Murakami}
\affil{Department of Physics, Faculty of Science, Ehime University,
    Matsuyama, 790-8577, Japan}
\email{\myemail}

\author{Karen M. Leighly, Chiho Matsumoto\altaffilmark{1}}
\affil{Department of Physics and Astronomy, The University of Oklahoma, 440 W. Brooks St., Norman, OK 73019}
\altaffiltext{1}{Current address: EcoTopia Science Institute,
Nagoya University, Furo-cho, Chikusa, Nagoya, 464-8603, Japan}

\author{Kiyoshi Hayashida}
\affil{Department of Astrophysics, Faculty of Science, Osaka University, Toyonaka, 560-0043,
Japan}

\and

\author{Dirk Grupe}
\affil{Astronomy Department, The Ohio State University, 140 W. 18th Avenue, Columbus, OH 43210}




\begin{abstract}
We present the results of timing analysis of the {\it XMM-Newton}
observation of the Seyfert 2 galaxy NGC 6300. The hard X-ray spectrum
above 2 keV consists of a Compton-thin-absorbed power law, as is often
seen in Seyfert 2 galaxies.  We clearly detected rapid time
variability on a time scale of about 1000 s from the light curve
above 2 keV. The excess variance of the time variability
($\sigma^{2}_{\rm RMS}$) is calculated to be $\sim$0.12, and 
the periodogram of the light curve is well represented by a power law 
function with a slope of 1.75. In contrast with previous results from 
Seyfert 2 nuclei, these
variability characteristics are consistent with those of Seyfert 1
galaxies.  
This consistency suggests that NGC 6300 has a similar black hole mass
and accretion properties as Seyfert 1 galaxies. Using the relation between
time variability and central black hole mass by Hayashida et al. (1998), 
the black hole mass of NGC 6300 is estimated to be $\sim2.8\ \times$ 
10$^{5}$ $M_{\odot}$.  Taking uncertainty of this method into account, 
the black hole mass is less than $10^{7}$ $M_{\odot}$. Taking the
bolometric luminosity of 3.3 $\times$ 10$^{43}$ erg s$^{-1}$ into 
consideration, this yields an accretion rate of $>$ 0.03 of the Eddington 
value, and comparable with estimates from Seyfert 1 galaxies using this 
method.  The time variability analysis
suggests that NGC 6300 actually has a Seyfert 1 nucleus obscured by a
thick matter, and more generally provides a new pillar of support for
the unified model of Seyfert galaxies based on obscuration.
\end{abstract}




\section{Introduction}
Active galactic nuclei (hereafter AGN) emit a huge amount of energy in
a wide energy band from radio to X-rays. It is generally thought that
the energy is generated by the release of gravitational energy of
matter accreted onto a central black hole.  AGN often show time
variability in continuum and line emission, and studies of the time
variability have led to advances in our understanding of the
structure of AGNs and the nature of nuclei themselves (e.g. Peterson et al. 
2000).  The study of
X-ray variability is one of the most promising endeavors to reveal the
characteristics of the nucleus, since X-rays are thought to originate
very close to the central black hole.  Early forays into the study of
time variability revealed a relationship between the variability time
scales and X-ray luminosities of Seyfert galaxies and quasars (e.g.,
Barr \& Mushotzky\ 1986; Nandra et al.\ 1997; Leighly\ 1999).  Fourier
analysis of the X-ray variability suggest characteristics of the
accretion disk (e.g., Pounds \& McHardy 1988, Kawaguchi et al. 2000).
Variability analysis has also been used to estimate the central black
hole mass (e.g., Hayashida et al. 1998 [hereafter H98], Markowitz et al. 2003).

Antonucci and Miller (1985) proposed the orientation-unified Seyfert
model, in which Seyfert 2 galaxies have Seyfert 1 nuclei obscured by
thick matter.  {\it Ginga} observations revealed the obscured nuclei
of Seyfert 2 galaxies in X-rays (e.g., Awaki et al.\ 1990; Awaki et al. 
1991), and {\it
ASCA} observations detected hard X-ray emission from many Seyfert 2
galaxies with $f_{\rm X} > 10^{-13}$ erg s$^{-1}$ cm$^{-2}$ in the
2--10 keV band (e.g., Turner et al.\ 1997).  Analysis of the X-ray
spectra showed clear-cut evidence that the nuclei of most Seyfert 2s
suffer significant absorption.  However, it remained unclear whether
the nuclei {\it beneath the obscuration} were really similar to those
of Seyfert 1 galaxies or not.

Variability analysis provides potentially a powerful probe to
understand the nature of the obscured nuclei, since, so long as the
obscuring column is Compton-thin, the pattern of the variability will
not be altered by the obscuration.  The Compton-thin Seyfert 2 galaxy 
NGC 5506, classified as a narrow emission line galaxy (NELG), showed a 
significant time variability, whose characteristics are similar to those 
of Seyfert 1 galaxies (e.g.,  Uttley, McHardy, \& Papadakis\ 2002 
[hereafter U02]).  However, the timing study of Seyfert 2 
galaxies except NELGs have not been well performed due to their faintness, 
and results so far, in fact, seem to
support a difference in the nuclei of Seyfert 1 and Seyfert 2
galaxies.  Turner et al. (1997) performed variability analysis on the
X-ray data of Seyfert 2 galaxies observed with {\it ASCA}. They
detected significant short time variability from NELGs. Most Seyfert 2 
galaxies, however, showed no significant variability.  
The lack of time variability within the 1-day {\it ASCA} observations may 
suggest that X-ray emissions of Seyfert 2 galaxies are dominated by 
scattering or Compton reflection.  If Seyfert 2 nuclei intrinsically show
the lack of the short-term variability, this result can be interpreted as 
evidence for rather massive black
holes with $M_{\rm BH} > 10^{7}$ $M_{\odot}$. Nishiura \& Taniguchi
(1998) derived the average dynamical masses of $10^{8} (1/\tau_{\rm
es})^{0.475}$ $M_{\odot}$ for nine Seyfert 2 galaxies using the
properties of the polarized broad H$\beta$ emission, where $\tau_{\rm
es}$ is the optical depth for electron scattering. For an optical
depth of 0.1, as suggested by Antonucci (1993), the dynamical masses of
Seyfert 2 nuclei are significantly larger than those of Seyfert 1
nuclei.

NGC 6300 is a nearby Seyfert 2 galaxy with a redshift of 0.0037
(Mathewson \& Ford 1996). This object was first detected in X-rays
serendipitously during a {\it Ginga} maneuver (Awaki 1991). A followup
{\it RXTE} observation revealed the flat spectrum and huge iron line
characteristic of Compton-reflection dominated Seyfert 2 galaxies
(Leighly et al. 1999).  But when observed using {\it Beppo-SAX}, the
spectrum revealed hard X-ray emission attenuated by a Compton-thin
absorber, suggesting that the central X-ray source had turned on.  The
{\it Beppo-SAX} observation also revealed rapid X-ray variability on a
time scale of a few $10^{3}$ s (Guainazzi 2002). However, a detailed
study of the time variability could not be carried out due to poor
statistics of the {\it Beppo-SAX} data.  {\it XMM-Newton} has the
largest effective area among imaging X-ray observatories, and the 
{\it XMM-Newton} observation of NGC 6300 provided time series data
of NGC 6300 with the highest quality.  It is possible to compare its
X-ray variability with those of Seyfert 1 galaxies, and to apply
Fourier analysis to the time series data.  We present the results of
the timing analysis, and discuss the characteristics of NGC 6300
nucleus in comparison with those of Seyfert 1 galaxies. The spectral 
analysis of the data will be presented in Matsumoto et al. (2004) 
(hereafter M04).

\section{Observations and Results}
NGC 6300 was observed by {\it XMM-Newton} on 2001-03-02 03:35 -- 16:35
(UT) for a total of 42 ks with the EPIC PN (Str\"uder et al. 2001)
and 46 ks with the EPIC MOS (Turner et al. 2001) detectors in the 
full-frame imaging mode using the medium filters.  We extracted the 
``flag=0'' events with the pattern of 0--4 and 0--12 from the PN and 
MOS detectors, respectively. 
We selected the events from the region within a 1$^{\prime}$ circle
centered on NGC 6300.  The X-ray spectrum could be described as a 
weak soft-component
plus a power-law component heavily absorbed by a thick matter with an
intrinsic column density $N_{\rm H}\ \approx$ 2.2 $\times$ 10$^{23}$
cm$^{-2}$ (M04). This spectral shape indicates that NGC 6300 was in
Compton-thin state.  Since the absorbed power law
component is considered to directly come from an obscured nucleus, 
we made the PN and MOS1+2 light curves with a bin size ($\Delta t_{\rm obs}$)
 of 256 s in the 2--10 keV band in order to study the obscured nucleus. 
Rapid variability with a large amplitude was seen in both the PN and 
MOS1+2 light curves in the hard band.  Since no background flares in a 
nearby blank field were seen, we are confident that we 
are observing variability from the nucleus.  The intensity dropped by a 
factor of two in only about 1000 s.
We summed up the PN data and the MOS1+2 data in the 2--10 keV band.
Figure 1 shows an X-ray light curve of NGC 6300 in the 2--10 keV band.
The values of average and mean-normalized excess variance ($\sigma^{2}_{\rm RMS}$) are 
0.83$\pm$0.01 c s$^{-1}$ and 0.121$\pm$0.014, respectively, 
where the error on  $\sigma^{2}_{\rm RMS}$ was computed according to 
Vaughan, Fabian, \& Nandra (2003). We also made light curves in the 0.2--2 keV band.
No significant time variability was found as mentioned by M04; a constant
model yields a reduced $\chi^{2}$ of 1.18 (163 dof). This result
suggests that the soft X-ray emitting region is not compact, and that
the soft component may originate from scattered light and/or diffuse
thermal emission, as is common among Seyfert 2 galaxies.
 
For characterizing the time variability of NGC 6300, we applied a slow 
Fourier transform to the PN+MOS light curve of NGC 6300. 
The power spectral density (PSD) is divided by the square of the mean
value ($\bar{s}$) of the source count rate ($s_{j}$) at time $t_{j}$.
The periodogram ($P(f)$ in units of Hz$^{-1}$) at frequency $f$ (Hz) 
is defined as

$\displaystyle P(f)=\frac{[a^{2}(f)+b^{2}(f)]T}{(\bar{s})^{2}},$

$\displaystyle a(f)=\frac{1}{n}\sum_{j=0}^{n-1}s_{j}\cos(2\pi ft_{j}),~~
b(f)=\frac{1}{n}\sum_{j=0}^{n-1}s_{j}\sin(2\pi ft_{j}),$

\noindent
where $T_{\rm obs}$ is the data length. 
The frequency range is about 2.4 $\times$ 10$^{-5}$ to 
3.9 $\times$ 10$^{-3}$ Hz, 
and there are about 80 frequency points. The periodogram is referred to as
normalized power spectral density (NPSD) in H98.
Raw periodograms usually have large scatter. Binning the periodogram in 
frequency or averaging multiple periodograms can reduce the scatter. 
Papadakis \& Lawrence (1993) pointed out that the binned logarithmic 
periodogram is a better estimator of the true underlying periodogram,
and that the errors of the binned logarithmic periodogram are similar to a
Gaussian distribution. The Gaussian distribution is essential to use
the $\chi^{2}$ statistics to estimate the goodness of a fit.
Therefore, we calculated the binned periodogram ($P_{\rm bin}(f_{\rm bin})$)
in logarithm-space,

$\displaystyle \log(P_{\rm bin}(f_{\rm bin})) = \frac{1}{M}\sum_{j}\log(P(f_{j})),$

$\displaystyle \log(f_{\rm bin}) = \frac{1}{M}\sum_{j}\log(f_{j}),$

\noindent
where $M$ is the number of bin in grouping the raw periodogram in frequency. 
We binned the raw periodogram with $M >$ 10, and obtained 7 points in the 
binned logarithmic periodogram (Figure\ 2).
We also estimated the Poisson noise of 1.2 Hz$^{-1}$.

Leighly et al. (1999) and Guainazzi (2002) found evidence for the
presence of a reflection component in the hard component with {\it
RXTE} and {\it Beppo-SAX}, respectively. M04 also find evidence for
such a component in the {\it XMM-Newton} observation, and estimate 
that the contribution of the reflection component to the hard component 
is about 15 \%.  Since the reflection component would be stable during one
day, the periodogram would be corrected for the stable reflection component
would have a normalization 1.4 times that shown in Figure\ 2.

\section{Estimation of Power-Spectral Shape}

We characterize the underlying shape of the observed periodogram in 
comparison with simulations, since we only obtained one measurement of 
the periodogram due to the relatively short observation.
U02 studied the errors on PSDs and the 
distorting effects of PSDs due to rebinning, red-noise leak and aliasing.
In order to take account of the errors and the distorting effects, a large
number of light curves should be simulated and they should be
analyzed in the same manner as the observed one.

We simulated 64 long light curves of length 16 $T_{\rm obs}$, since 
U02 mentioned that simulated light 
curves must be significantly longer than the observed light curve for 
taking account of red-noise leak. For the time resolution 
$\Delta t_{\rm sim}$ of simulated light curves, we set to be 
$\Delta t_{\rm sim}$ = $\frac{1}{16} \Delta t_{\rm obs}$ for taking account 
of aliasing effect.
The long light curves are simulated using the method by Timmer 
\& Ko\"nig (1995) in assuming that a power spectral density is described 
with a power law function for frequencies $>$ 2$\times10^{-5}$ Hz and a 
constant for lower frequencies, since the PSD shape 
for AGNs could be described well by a power law function of frequency; 
i.e., $P(f) \propto f^{-\alpha}$, with a break at low frequencies in 
general (e.g. Lawrence \& Papadakis 1993; Markowitz \& Edelson 2001). 
The normalization of the PSD is determined by using the observed excess 
variance of 0.121.

We split each long light curve into 16 segments, then we obtained 1024 
light curves with length $T_{\rm obs}$. All light curves were binned 
with the bin size $\Delta t_{\rm obs}$. The 
averaged values of the mean count rates and the excess variances of 
the 1024 simulated light curves become equal to the observed values
($\bar{s}$ and $\sigma^{2}_{\rm RMS}$).
Figure\ 3 shows the distributions of the mean count rates and excess 
variances of the 1024 light curves simulated using the slope of 1.75.
From these distributions, the standard deviations of the mean count rate
and the logarithmic excess variance are calcurated to be 0.15 and 0.25,
respectively. The standard deviation of the logarithmic excess variance corresponds
to $^{+0.08}_{-0.05}$ in a linear scale.  These standard deviations are larger than 
errors on the observed values. This is caused by the stochaostic nature 
of the mean value and the excess variance. These values measured from a single light 
curve show a large scatter around the underlying value.

Suitable Poisson noise was added to the simulated 
light curves, and periodograms were extracted and rebinned as for the 
real one. We then constructed an average periodogram
($\overline{P_{\rm sim}}(f_{\rm i})$) with errors 
($\sigma_{\rm sim}(f_{\rm i})$) by obtaining the standard deviation at 
each rebinned frequency point ($f_{\rm i}$).  
Please note that the averages and errors are calculated in logarithm-space.
We compared them with the real data ($P_{\rm obs}(f_{\rm i})$) using the 
$\chi^2$ method,

$\displaystyle \chi^{2}=\sum_{i}\frac{[\overline{P_{\rm sim}}(f_{\rm i})-P_{\rm obs}(f_{\rm i})]^2}{\sigma_{\rm sim}^{2}(f_{\rm i})}.$

\noindent
We calculated a $\chi^2$ value at a given power law slope $\alpha$, then 
we found the best fit slope of 
1.75$^{+0.25}_{-0.15}$ with the minimum $\chi^2$, where the error of 
the slope was determined by $\Delta \chi^2=1$ (Table\ 1).
Figure 4 shows the averaged periodogram for the best fit slope.

\section{Discussion}
\subsection{Characteristics of Time Variability}

We report rapid time variability in the {\it XMM-Newton} light curve
of the Seyfert 2 galaxy NGC 6300. Its $\sigma^{2}_{\rm RMS}$ is about
0.121. Comparison with the excess
variances from {\it ASCA} observations of Seyfert 1 galaxies reported
by Nandra et al.\ (1997) shows that NGC 6300 is one of the most
variable sources among Seyfert galaxies. The intrinsic luminosity 
(i.e., after accounting for the Compton-thin absorption) in the 2--10 
keV band is estimated to be 1.2 $\times$ 10$^{42}$ erg s$^{-1}$ 
assuming $H_{\rm 0}$ = 50 km s$^{-1}$ Mpc$^{-1}$ (M04). NGC 6300 
lies on the $\sigma^{2}_{\rm RMS}$--$L_{\rm X}$ anticorrelation found 
by Nandra et al.\ (1997) for Seyfert 1 galaxies (see Figure\ 5). 
Turner et al. (1997) performed timing analysis for Seyfert 2 galaxies
observed with ASCA. They found that three bright sources (NGC 526A, 
MCG --5-23-16, and NGC 7314) among five bright galaxies were variable, 
and that their 
excess variances were consistent with those of Seyfert 1 galaxies.  
Since these three galaxies are classified as X-ray selected narrow emission
line galaxies (NELG), NGC 6300 is the first Seyfert 2 galaxy with a rapid
time variability except NELG.

The power spectral density is well represented by a power law function 
with $\alpha\ \approx\ $1.75.  The slopes of the periodograms of Seyfert 
1 galaxies are distributed between 1--2 over the frequency range of 
$\sim$10$^{-5}$ to $\sim$10$^{-3}$ Hz; (e.g. Lawrence \& Papadakis 1993).
Figure 6 shows a distribution of the slopes for Seyfert 1 galaxies.
The broad-band periodograms of Seyfert 1 galaxies can be well represented 
by a knee model (Lawrence \& Papadakis 1993) or a break model (U02).  We
therefore chose the slope in the higher frequency range ($> 10^{-5}$Hz) for 
comparison with our result. We found that the slope of NGC 6300 is 
similar to those of Seyfert 1 galaxies.

Recent analysis suggest that the $f^{-\alpha}$ 
fluctuations are exhibited by a magnetized accretion disk. 
Kawaguchi et al. (2000) suggest that the slope $\alpha$ is characterized 
by the distribution of cluster size in an accretion disk, and Merloni \&
Fabian (2001) suggest that the slope $\alpha$ is associated with the 
amplitude of the avalanches in their 'thunder cloud' model. 
The similarity of the slope $\alpha$ and the similarity of the 
$\sigma^{2}_{\rm RMS}$--$L_{\rm X}$ relation would indicate that 
NGC 6300 actually has a similar central engine with Seyfert 1 galaxy.

\subsection{Estimation of Central Black Hole Mass}

The black hole mass could be estimated using a PSD, because 
the variability time scale is associated with the size of emitting region.
H98 introduced a characteristic frequency of the
variability time scale, and estimated the central black hole mass.
U02 estimated the black hole mass using
the frequency of the break of a periodogram. We try to estimate the central 
black hole mass from the periodogram of NGC 6300.
Although U02 pointed out that a break frequency is useful to
estimate the black hole mass, we can not well deduce the
break frequency of NGC 6300 due to poor statistics and limited observation
time.  Thus, following the procedure discussed in H98, we determined a
characteristic frequency, $f_{\rm 0}$, defined by $f_{\rm 0}\ \times\ $
P($f_{\rm 0})\ =\ 10^{-3}$, and then used that to estimate a central
black hole mass in NGC 6300.  

We obtained the characteristic frequency $f_{\rm 0}$ of
$1.6^{+6.4}_{-0.8}\ \times\ 10^{-3}$ Hz at 1 $\sigma$ confidence. 
Using the formula $M_{\rm BH}$ = 10 $\times$ [45.5/$f_{\rm 0}$ (Hz)] 
$M_{\odot}$
by H98, we deduced a central black hole mass of
$2.8^{+2.9}_{-2.2}\ \times\ 10^{5}\ M_{\odot}$.
We note that Poisson noise is significant at $f\ =\ 1.6\ \times$ 10$^{-3}$ Hz,
and that we estimate the characteristic frequency assuming that we can
simply extrapolate the periodogram to higher frequency. 
The extrapolation might become a cause of uncertainty of our estimation.
Markowitz et al. (2003) claimed that the characteristic 
frequency depends on the PSD break frequency and high frequency slope. 
However, the stellar black hole candidate Cyg X-1 shows differing PSD 
shapes with differing break frequencies in its high and low state, but shows 
similar PSD shapes in the high frequency part (Belloni \& Hasinger 1990).  
Despite the change in PSD spectral state, $f_0$ for Cyg X-1 is relatively stable.
For instance, $f_{\rm 0}$ is to be about 30 Hz for the high/soft state of Cyg X-1 and
100 Hz for the low/hard state of Cyg X-1 (Done \& Giertli\'{n}ski 2005). However, Done \&  
Giertli\'{n}ski (2005) pointed out that Cyg X-1's timing properties are
not typical of galactic black hole candidates. Thus, we have to consider the difference of 
the timing properties for different sources as well as in different states. The 
ambiguity of the timing properties may lead to uncertainty of $f_{\rm 0} - M_{\rm BH}$  
relation.
H98 pointed out that the masses derived in this method are smaller by 1 or 2 
orders of magnitude than those estimated by other methods, e.g. emission line 
width.  The black hole mass of NGC 6300 would be conservatively
estimated to be less than 10$^{7}$ $M_{\odot}$.

The method of Nikolajuk, Papadakis \& Czerny (2004) can also be 
used to yield an estimate of NGC 6300's black hole mass from the 
excess variance measurement. Their equation 4 is 
$M_{\rm BH}$ = $C$ ($T_{\rm obs}$ - 2 $\Delta t$) / $\sigma^2_{\rm rms}$,
where $C$ is a constant, estimated to be 0.96 $\pm$ 0.02, based on 
excess variance measurements in Cyg X-1, $T_{\rm obs}$ is the light curve 
duration (42 ks in this case),  $\Delta t$ is the sampling time 
(256 s), and $\sigma^2_{\rm rms}$ is the mean-normalized excess variance.  
If we assume that the {\it XMM-Newton} light curve is probing 
variations above the break frequency in NGC 6300 (i.e., if the break 
is less than 1/42000 s = 2.4 $\times 10^{-5}$ Hz), then, given the 
excess variance measurement of 0.121$^{+0.08}_{-0.05}$, the black hole mass 
would be 3.3$^{+2.4}_{-1.4}~\times 10^5$ $M_{\odot}$. This is very similar to 
the mass estimate obtained using the H98 method. We note, however, 
that any uncertainty in the excess variance translates into an 
uncertainty in black hole mass estimate, and an average of multiple 
excess variance measurements, if an average existed, would yield a 
more "stable" black hole mass estimate compared to that obtained from 
a single excess variance measurement.

We compared our estimate of the black hole mass with estimates
obtained using other methods.  One method is to use the tight
relationship between central black hole masses and velocity
dispersions (e.g., Gebhardt et al. 2000; Merritt \& Ferrarese 2001).
Buta (1987) estimated a central velocity dispersion $\sigma_{v}$ of 113
km s$^{-1}$ from his analysis of a NGC 6300 rotation curve, although
this value is slightly smaller than the estimate from the
Faber-Jackson law ($L\propto \sigma^{4}_{v}$) (143 km s$^{-1}$).
Using the velocity dispersion of 113 km s$^{-1}$, and the $\sigma_{v}$--
$M_{\rm BH}$ relation by Merritt \& Ferrarese (2001), we find a
central black hole mass estimate of $\sim$10$^{7}$ $M_{\odot}$, which
is somewhat larger than the estimate obtained by the X-ray timing
analysis.  Since Buta (1987) pointed out that the rotation curve was
poorly defined in the central region of NGC 6300, this discrepancy may 
be caused by an uncertainty of the central velocity dispersion
estimated from the rotation curve. Detailed observations of $\sigma_{v}$ 
may improve the estimate of the central black hole mass.
Note that this black hole mass estimate of 10$^7 M_{\odot}$ is not much 
greater than the black hole mass of MCG--6-30-15, estimated to be
$M_{\rm BH}\ \approx\ 3\times10^{6}\ M_{\odot}$ based on PSD analysis and a stellar 
velocity dispersion measurement of $\sigma_{v} \approx\ 94\pm 9$ km s$^{-1}$
(McHardy €et €al. 2005).  The break in the X-ray PSD of MCG--6-30-15
is near $8 \times 10^{-5}$ Hz (McHardy et al. 2005); one might therefore
expect a break in the PSD of NGC 6300 near this frequency or just 
slightly lower. Unfortunately, the current data cannot constrain any 
such break in the PSD, as noted above.

It has been observed that the width of the [\ion{O}{3}] line is
correlated with the stellar velocity dispersion (Nelson \& Whittle
1996).  This correlation has been shown to carry through to the black
hole mass (Nelson 2000).  Although this correlation is not as tight as 
that between the black hole mass and the stellar velocity dispersion,
it can be used to provide another estimate of the black hole masses in
our objects.  
Given that the black hole masses of NGC 6300 and MCG--6-30-15
might not be too vastly different (see above),
we use the [\ion{O}{3}] line 
profile from MCG--6-30-15 for purposes of comparison.
The spectra for both objects were obtained using the CTIO Blanco 4-meter
telescope and the R-C spectrograph, in June 1999 and June 1998
respectively, using a $2^{\prime\prime}$ slit with typical seeing of
1--$1.5^{\prime\prime}$. The resolution in both cases measured from
the calibration lamp data was $4.4$ \AA\/, and the lines were resolved,
although close to the resolution limit of the data. The [\ion{O}{3}]
lines were fit with a Gaussian model; a second Gaussian was used to
model the small blue wing in NGC 6300.  After accounting for the
resolution, we obtain a velocity FWHM of $230\ \rm km\, s^{-1}$ and
$180\ \rm km\, s^{-1}$ for NGC 6300 and MCG--6-30-15.  These values are
confirmed by high resolution spectroscopy reported in the literature.
Whittle (1992) list [\ion{O}{3}] FWHM of $220\ \rm \, km\,s^{-1}$ from
high signal-to-noise  spectra with resolution better than 2 \AA\/ for
NGC 6300.  Busko \&  Steiner (1988) report a FWHM of [\ion{O}{3}] of
$140 \pm 3 \rm \,km\,s^{-1}$ with an instrumental resolution of $35\ \rm km\,
s^{-1}$ in MCG--6-30-15.  They comment that MCG--6-30-15 lies on the
extreme of the FWHM distribution in their sample of AGN.  Using the
regression obtained by Nelson (2000) (his Fig.\ 1), we estimate
$log(M_{\rm BH})$ of around 6.5--6.6, intermediate between the
estimates obtained from the time series analysis and velocity
dispersion in NGC 6300.  Again, we acknowledge that there is quite a
large dispersion in the [\ion{O}{3}] method for estimating the black
hole mass (Nelson (2000) quote a mean error of 0.4 dex); however, we
emphasize that both NGC 6300 and MCG--6-30-15 have extremely narrow
[\ion{O}{3}] lines compared with the Nelson (2000) sample.  

In order to characterize the mass accretion, we computed the Eddington
ratio, which is the ratio between the bolometric and Eddington
luminosities, for NGC 6300.  The intrinsic X-ray luminosity in the
2--10 keV band was 1.2 $\times$ 10$^{42}$ erg s$^{-1}$. For the 
bolometric correction of 27.2 given by Padovani \& Rafanelli (1988), the
bolometric luminosity was estimated to be 3.3 $\times$ 10$^{43}$ erg
s$^{-1}$, which leads to the mass-accretion rate of $\sim$5.6 $\times$
10$^{-3}$ $M_{\odot}$ yr$^{-1}$ for a matter-to-energy conversion of
0.1. The Eddington luminosity for 2.8 $\times$ 10$^{5}$ $M_{\odot}$
was 3.5 $\times$ 10$^{43}$ erg s$^{-1}$. The Eddington ratio is
inferred to be 0.9. Considering the uncertainty of the central black hole
mass, the Eddington ratio would be in the range from $\sim$0.03 to 
$\sim$0.9, which is also similar to those values obtained
from Seyfert 1 galaxies (e.g. H98).

\section{Conclusion}

We detected short time scale variability of the Seyfert 2 galaxy 
NGC 6300 on the order of $\sim1000\,\rm s$ using {\it XMM-Newton}.  
The $\sigma^{2}_{\rm RMS}$ was estimated to be 0.121 and the 
periodogram could be well represented by a power-law function with
a slope of 1.75. These characteristics are similar to those of Seyfert 1
galaxies. We estimated the central black hole mass in NGC 6300 at 
$\sim2.8\ \times\ 10^{5}$ $M_{\odot}$ based on the scaling relation 
between X-ray variability timescales and black hole masses. Considering a
scatter of the relation, the central black hole mass is less than
$10^{7}\ M_{\odot}$. This is consistent with the estimates by other methods
($10^{7}\ M_{\odot}$ by $\sigma_{v}-M_{\rm BH}$ relation and $10^{6.5-6.6}
\ M_{\odot}$ by the width of [\ion{O}{3}]).
We also estimated the Eddington ratio of 0.1--0.9 from the bolometric 
luminosity of 3.3 $\times$ 10$^{43}$ erg s$^{-1}$.
These results point to similar physical conditions (accretion disk,
black hole mass) in the central engines of NGC 6300.

Thus, from the X-ray time series analysis of the Seyfert 2 galaxy NGC
6300, and comparison with results from time series analysis of Seyfert
1 galaxies, we conclude that NGC 6300 has a Seyfert 1 nucleus obscured
by thick matter. This result supports the unified Seyfert theory.
Furthermore, our result demonstrates that {\it XMM-Newton} is useful
for studying the time variability of Seyfert 2 galaxies, and may
provide a new tool to reveal the characteristics of Seyfert 2 nuclei.
 
\acknowledgments
The authors wish to thank for the staff of {\it XMM-Newton} data center,
and the anonymous referee for thoughtful comments and suggestions.
 This study was carried out in part by the
Grant-in-Aid for Scientific Research (14340061 \& 17030007) of the 
Ministry of Education, Culture, Sports, Science and Technology (KH \& HA).



\clearpage


\begin{figure}
\plotone{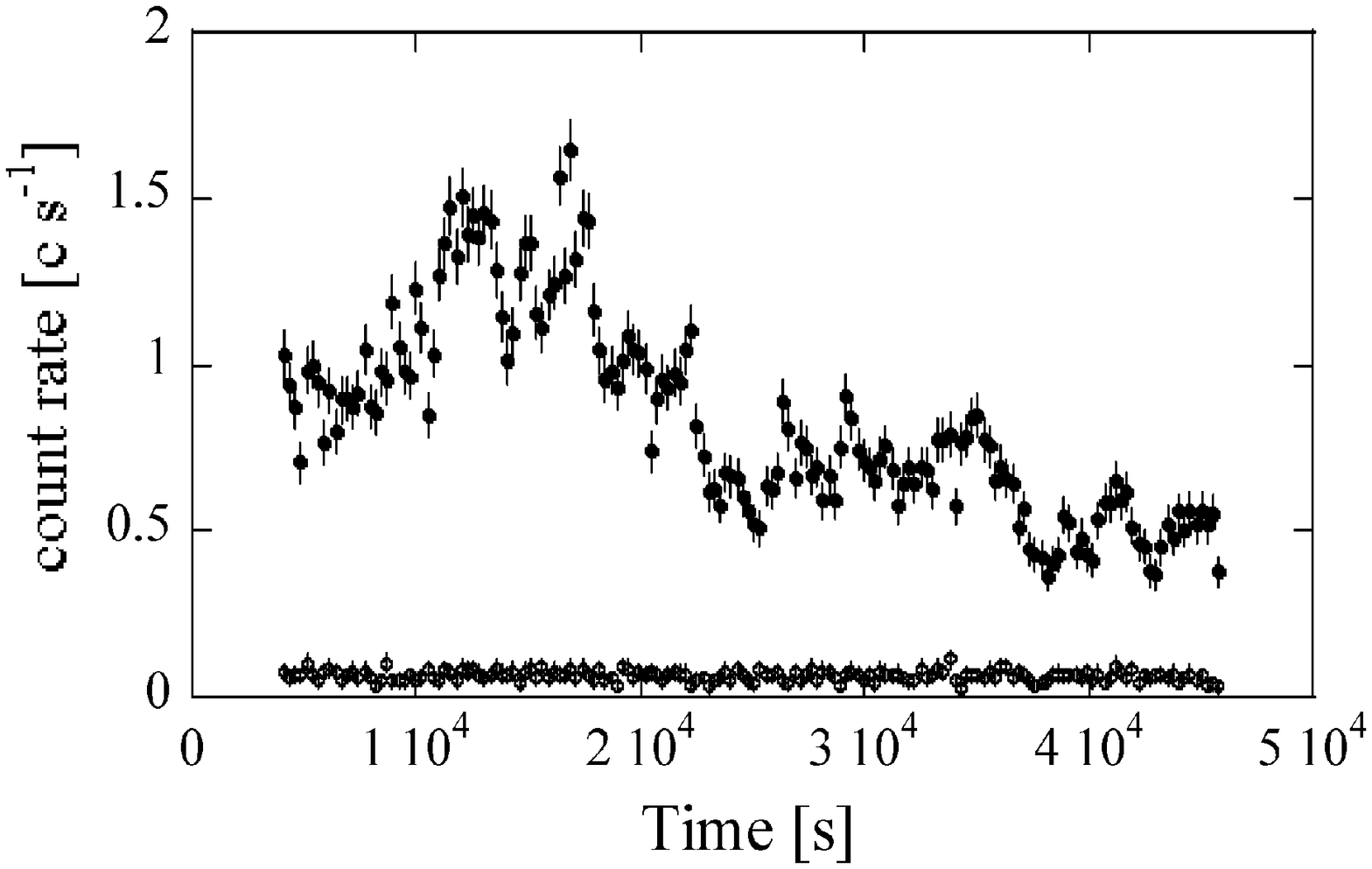}
\caption{X-ray light curves of NGC 6300 in the 0.2--2 keV (lower in
  the panel) and 2--10 keV (upper in the panel) bands
with the XMM-Newton. The MOS1+2 and PN data were summed up. The binning 
time is 256 s.  The 2--10 keV light curves show large-amplitude rapid 
variabilities with excess variance of $\sim$0.12, while the 0.2--2 keV 
light curves show no significant variability. 
\label{fig1}}
\end{figure}

\begin{figure}
\plotone{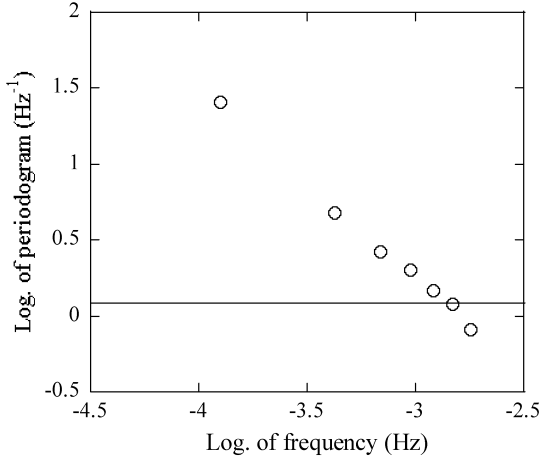}
\caption{A periodogram of the PN+MOS12 (2--10 keV) light 
curve of NGC 6300 (open circles).  
The solid line displays the Poisson noise level of 1.2 Hz$^{-1}$.
\label{fig2}}
\end{figure}

\begin{figure}
\plottwo{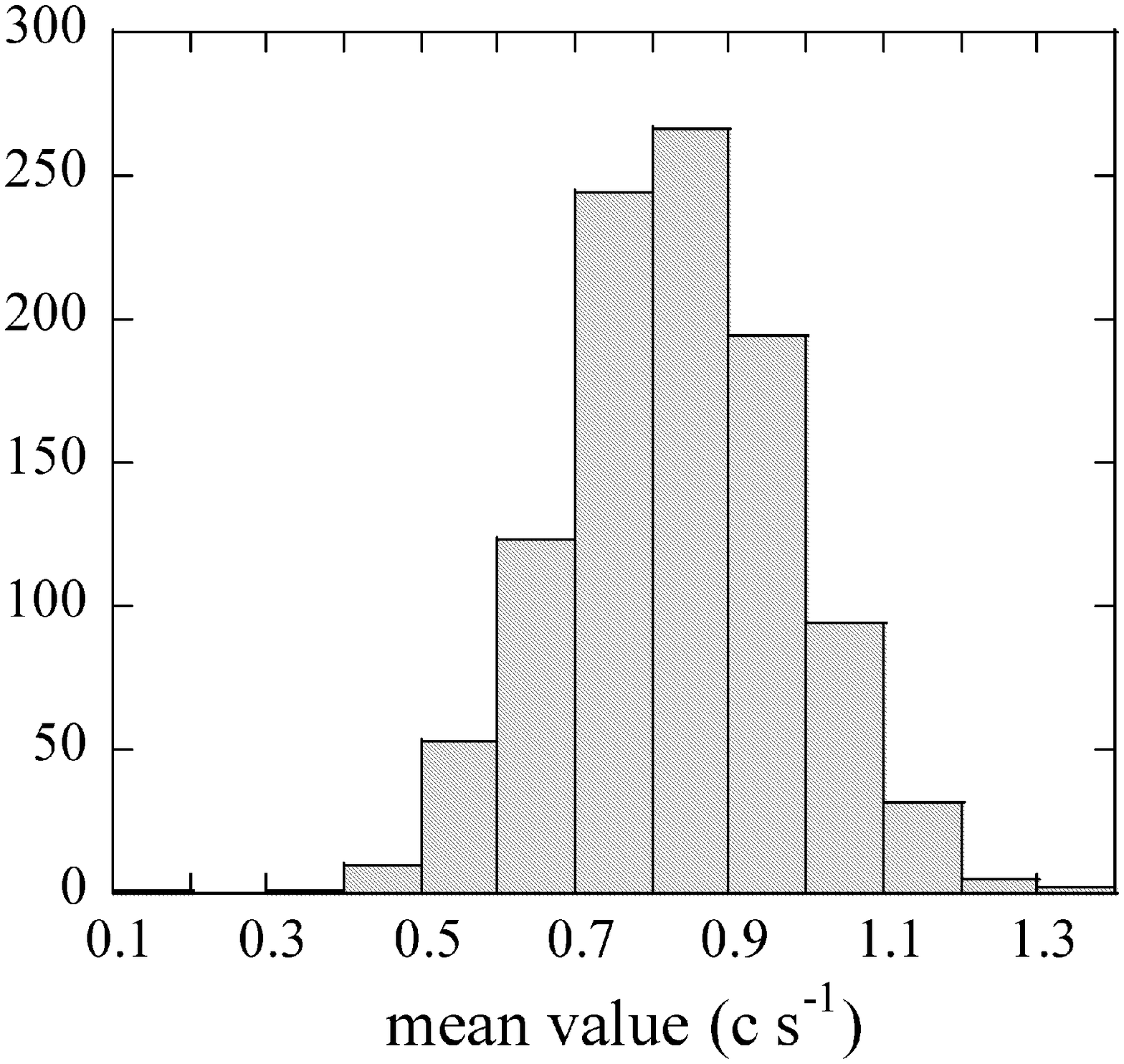}{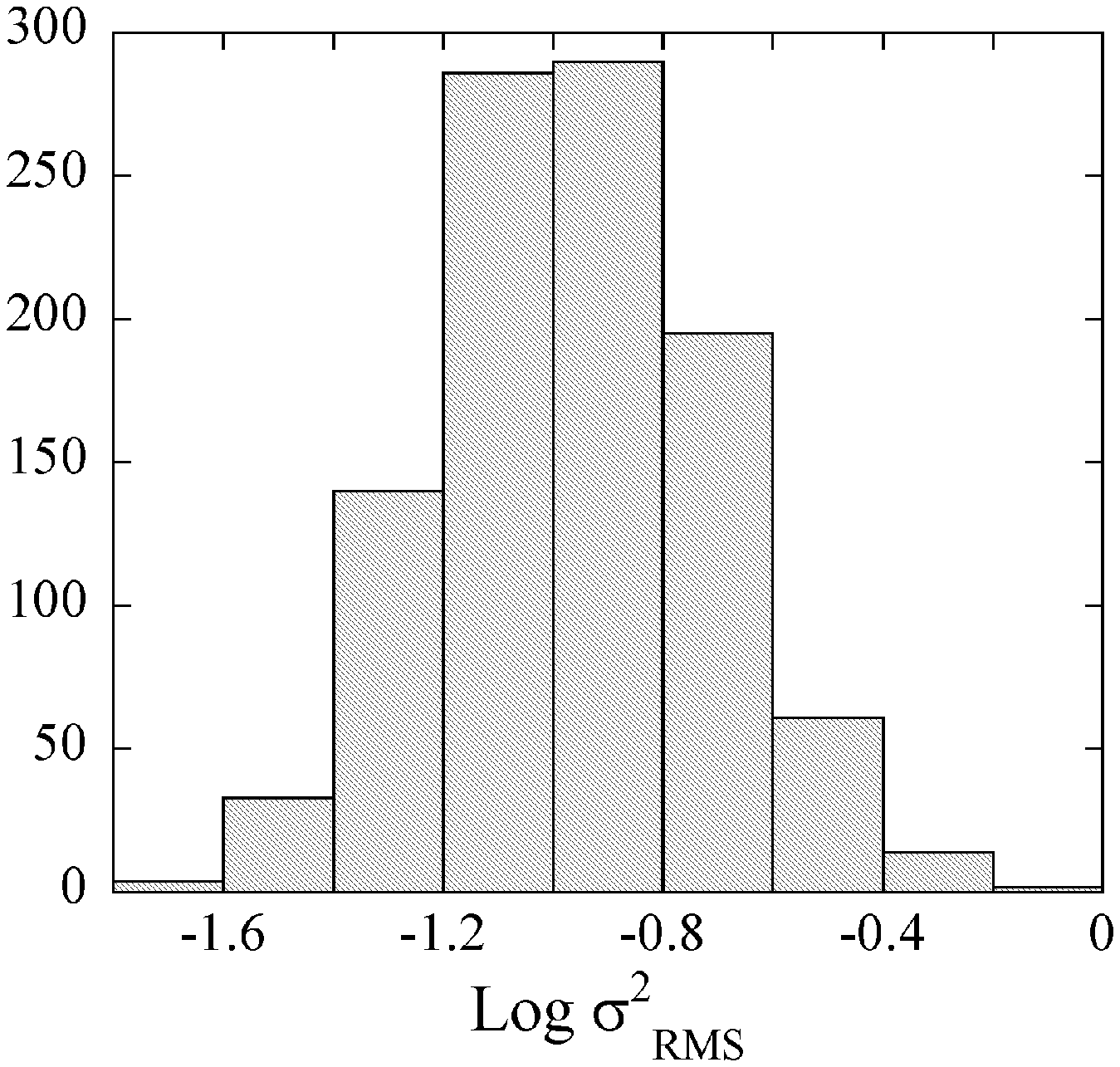}
\caption{The distributions of the mean count rate (left) and logarithmic 
excess variance (right).  The 1024 light curves are simulated  with
a power law slope of 1.75.  
\label{fig3}}
\end{figure}

\begin{figure}
\plotone{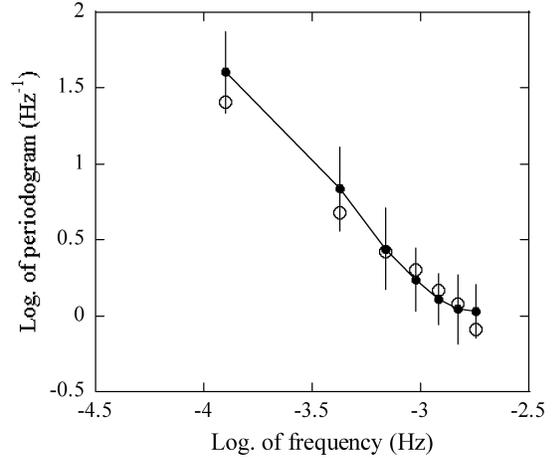}
\caption{The average of the 1024 simulated periodogram with the best fit 
slope of 1.75. 
The errors are derived by the deviation of the 1024 simulated periodograms.
\label{fig4}}
\end{figure}

\begin{figure}
\plotone{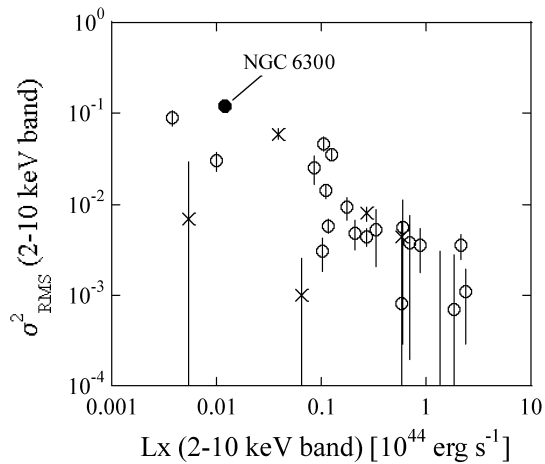}
\caption{Excess variance of NGC 6300 overlaid on those of Seyfert sample.
NGC 6300 is represented by the filled circle, and the Seyfert 1 sample
taken from Nandra et al. (1997) are open circles.  The Seyfert 2 and NELG 
sample taken from Turner et al. (1997) are crosses.
\label{fig5}}
\end{figure}

\begin{figure}
\plotone{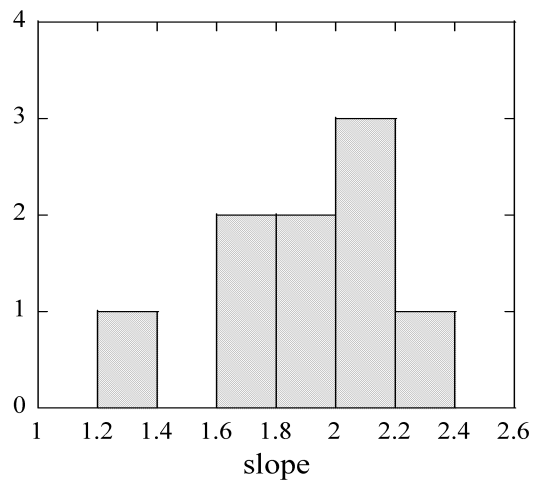}
\caption{A distribution of the slopes of Seyfert 1 galaxies. 
We chose the slope in the higher frequency range ($> 10^{-5}$ Hz) 
from Lawrence \& Papadakis (1993) and U02.
\label{fig6}}
\end{figure}

\clearpage






\clearpage
\begin{deluxetable}{cccc}
\tabletypesize{\scriptsize}
\tablecaption{Result from the fit of the periodogram with $f^{-\alpha}$ model \label{tbl-1}}
\tablewidth{0pt}
\tablehead{
\colhead{slope} & \colhead{$f_{\rm 0}$} & 
\colhead{$\chi^{2}$/dof }  & \colhead{$M_{\rm BH}$} 
\\
\colhead{  } & \colhead{ ($\times\ 10^{-3}$ Hz)}   & \colhead{} & 
\colhead{ ($\times\ 10^{5} M_{\odot}$)}
}
\startdata
1.75$^{+0.25}_{-0.15}$  & 1.6$^{+6.4}_{-0.8}$  & 1.2/5 & 2.8$^{+2.9}_{-2.2}$ \\
\enddata






\tablecomments{Errors on the model parameters correspond to 1$\sigma$ confidence level for one interesting parameter.}
\end{deluxetable}









\end{document}